\documentclass[12pt,preprint]{aastex}

\usepackage{rotating}

\usepackage{booktabs,threeparttable}

\slugcomment{To appear in Apj}
\shorttitle{Dark Matter Fraction in Lens Galaxies: New Estimates from Microlensing}
\shortauthors{JIM\'ENEZ-VICENTE ET AL.}

\begin{document}

\title{Dark Matter Mass Fraction in Lens Galaxies: New Estimates from Microlensing}

\author{J. JIM\'ENEZ-VICENTE\altaffilmark{1,2},  E. MEDIAVILLA\altaffilmark{3,4},
C. S. KOCHANEK\altaffilmark{5},  J. A. MU\~NOZ\altaffilmark{6}}

\altaffiltext{1}{Departamento de F\'{\i}sica Te\'orica y del Cosmos, Universidad de Granada, Campus de Fuentenueva, 18071 Granada, Spain}
\altaffiltext{2}{Instituto Carlos I de F\'{\i}sica Te\'orica y Computacional, Universidad de Granada, 18071 Granada, Spain}
\altaffiltext{3}{Instituto de Astrof\'{\i}sica de Canarias, V\'{\i}a L\'actea S/N, La Laguna 38200, Tenerife, Spain}
\altaffiltext{4}{Departamento de Astrof\'{\i}sica, Universidad de la Laguna, La Laguna 38200, Tenerife, Spain}
\altaffiltext{5}{Department of Astronomy and the Center for Cosmology and Astroparticle Physics, The Ohio State University, 4055 McPherson Lab, 140 West 18th Avenue, Columbus, OH, 43221 }
\altaffiltext{6}{Departamento de Astronom\'{\i}a y Astrof\'{\i}sica, Universidad de Valencia, 46100 Burjassot, Valencia, Spain.}

\begin{abstract} We present a joint estimate of the stellar/dark matter mass fraction in lens galaxies
and the average size of the accretion disk of lensed quasars from microlensing measurements of 27 quasar image pairs
seen through 19 lens galaxies. The Bayesian estimate for the fraction
of the surface mass density in the form of stars is $\alpha=0.21\pm0.14$ 
near the Einstein radius of the lenses ($\sim 1 - 2$ effective
radii). 
The estimate for the average accretion disk size is $R_{1/2}=7.9^{+3.8}_{-2.6}\sqrt{M/0.3M_\sun}$ light days. 
The fraction of mass
in stars at these radii is significantly larger than previous estimates from 
microlensing studies assuming quasars were point-like.
The corresponding local dark matter
fraction of 79\% 
is in good agreement with other estimates based on strong lensing or 
kinematics.   
The size of the accretion disk inferred in the present study is
slightly larger than previous estimates. 

\end{abstract}


\section{Introduction}

The amount and distribution of dark matter relative to stars
is a crucial probe of early-type galaxy structure. 
In particular, changes in the dark matter fraction
with radius provide important 
information about the mechanisms of galaxy formation and the interaction
of dark and baryonic matter during the initial collapse 
(including processes like baryonic 
cooling, settling, star formation and feedback) and subsequent mergers (see Diemand
\& Moore 2011 for a review). 

But measuring this dark matter fraction 
is difficult. 
Existing estimates use X-ray observations, 
stellar dynamics or gravitational lensing, and each of these methods
has its own advantages and difficulties.
X-ray observations of the hot gas in massive galaxies can provide
an estimate of the total mass under the assumption of 
hydrostatic equilibrium (see Buote \& Humphrey, 2012).
This method is very robust and simple, with its main uncertainties
coming from the robustness of the hydrostatic equilibrium hypothesis,
the possibility of
non-thermal contributions to the pressure and contamination from
emission by
larger scale group/cluster halo gas.
Stellar dynamics can also be used to estimate the structure of
the gravitational potential (see for example Courteau et al. 2014). 
In this case, the structure of the orbits (anisotropy) is the primary
source of uncertainty. With both X-ray and stellar dynamics it is difficult
to extend the measurements to large radii.

Gravitational lensing is also a very powerful
probe of dark matter because it provides
direct measurements of the total mass of the system (within a certain
radius) regardless of whether it is dark
or baryonic. On large scales, weak lensing can be used to estimate 
the mass distribution in the outer parts of halos. Such studies
have shown that the mass profiles at those radii 
are consistent with the Navarro, Frenk \& White (1997) (hereafter NFW) or Einasto (1965) profiles predicted
by simulations 
(e.g. Mandelbaum, Seljak \& Irata 2008). 
The inner regions of galaxies are
more complex, as 
baryons influence the mass profile
and can make the halos significantly steeper (Blumenthal et al. 1986, Gnedin et al. 2004). In these inner regions, strong lensing can be used 
to robustly estimate the total galaxy mass
within the Einstein Radius of the lens (typically 1-2 effective radii).
Indeed, this estimate of the total projected mass 
inside the Einstein radius of the lens galaxy is very robust, 
and depends very weakly on the specific lens model 
(i.e. on the specific radial profile or the angular structure of the lens)
and this can be used to statistically constrain the structure of
galaxies (Rusin \& Kochanek 2005).
The radial mass distribution can be constrained if additional or extended
images exist (e.g. Sonnenfeld et al. 2012), or by combining lensing with stellar dynamics (e.g. Romanowsky \& Kochanek 1999, Koopmans et al. 2006).
However, dividing the measured mass between dark matter and stars
is more difficult, as it requires a model of the stellar mass. 
Photometry, in combination with stellar population synthesis (that can
provide an estimate for the stellar M/L ratio)
can be used to estimate the stellar mass distribution 
(see for example Jiang \& Kochanek 2007, Auger et al. 2009, Tortora et al. 2010, Leier et al. 2011, Oguri et al. 2014). 
Nevertheless, in this procedure there
is always a great uncertainty due to the IMF of the stars, particularly
given recent arguments in favour of 
``bottom heavy'' and variable IMFs (van Dokkum \& Conroy 2010, 2011; Conroy \& van Dokkum 2012).
Examples of lensing studies are Rusin \& Kochanek (2005), Koopmans et al. (2006),
Auger et al. (2010), Leier et al. (2011) and Oguri et al. (2014), 
and these studies generally find that the integrated dark matter fraction
inside the Einstein radius is roughly 0.3-0.7. 
Estimates of the local value at the Einstein radius are more model dependent.

Microlensing of the images of gravitationally lensed quasars 
provides a direct means of measuring the dark matter fraction at the location of
the lensed images.
Microlensing is caused by the granularities in the mass distribution 
created by stars and their remnants 
which induce time dependent changes in the flux of the 
lensed quasar images (see the review by Wambsganss, 2006). 
At any instant, they produce flux ratio anomalies 
that cannot be accounted for
by the smooth macro model of the lens. 
Particularly when the stars are only a small fraction of the surface mass
density, microlensing is very sensitive to the relative fractions
of stars and dark matter near the
images (e.g. Schechter \& Wambsganss, 2004). 
We can therefore estimate the local fraction of mass in stars or
dark matter from the statistics of microlensing.
Recently, this effect has also been used to calibrate the stellar mass
fundamental plane by Schechter et al. (2014).

The main practical difficulty of this method resides in the 
determination of the flux anomalies generated by microlensing. 
The problem is that the anomalies are usually 
identified assuming that a standard 
macrolens model 
can be used as an absolute ``flux ratio'' reference, without
contamination by differential extinction 
(e.g. Falco et al. 1999, Mu\~noz et al. 2004) or perturbations from 
substructures in the lens (e.g. Dalal \& Kochanek 2002, Keeton et al. 2003).
One solution is to use 
emission line ratios, which are little affected by microlensing (e.g., Guerras et al. 2013), as a baseline 
to remove the effects of the macro-magnification, extinction and substructure
(Mediavilla et al. 2009, hereafter MED09).
There remains the problem of intrinsic source variability modulated
by the lens time delays
as a contribution of apparent flux anomalies.
At optical wavelengths, the amplitude of quasar variability on 
timescales of the order of the
time delays between images is rather modest, and 
should be a source of some extra noise rather than a significant bias on
the results (see the discussion in MED09).  

To date, there have been two microlensing studies of 
the stellar mass fraction using ensembles of lenses, and they
obtained similar values for the 
fraction of mass in 
stars. MED09, using optical flux ratios of 29 images in
20 lenses, found a stellar 
surface density of
5\%, and Pooley et al. (2012), using X-ray flux ratios of images in 14 lenses, 
found a fraction of 7\% near
the Einstein radius of the lenses. 
A third study based on microlensing for only three lenses by Bate et al. (2011) 
found much higher values, in the range 20\% to 100\%.  
There are also several microlensing results for individual lenses that 
usually favour dark matter
dominated galaxies with stellar fractions at the image positions 
roughly in the range 8-25\% 
(Keeton et al. 2006, Kochanek et al. 2006, 
Morgan et al. 2008, Chartas et al. 2009, Pooley et al. 2009, Dai et al. 2010b, Morgan et al. 2012).
The exception to this rule is the lens Q2237+0305, where microlensing is 
dominated by bulge stars and is therefore compatible with nearly 100\% of the
surface mass density in form of stars (Kochanek 2004, Bate et al. 2011, Pooley et al. 2012). These estimates are somewhat larger than the microlensing estimates from lens samples by MED09 and Pooley et al. (2012).
While it is not straightforward to compare local and 
integrated values for the stellar mass fractions, there also 
seems to be an apparent discrepancy 
between the low local values of 0.05-0.07 determined by the microlensing samples and 
the (integrated) values of 0.3-0.7 estimated by other means.

The studies of microlensing in individual lenses are largely focussed on
measuring the sizes of quasar emission regions, with the dark matter fraction
as a ``nuisance'' parameter, so
the source size is included as an unknown
in the calculations, while the two large statistical microlensing studies (MED09, Pooley et 
al. 2012) were 
done under the 
hypothesis that the size of the source
is very small compared to the Einstein radii of the microlenses.
At the time of these studies, it seemed plausible that source sizes were
small enough to be neglected, although MED09 did point out that
there was a clear covariance in the sense 
that larger source sizes lead to larger stellar mass fractions.
Recent estimates of quasar accretion disk sizes 
(see, e.g., Morgan et al. 2008, 2010, 2012, Blackburne et al. 2011, Mediavilla et al. 2011b, Mu\~noz et al. 2011, Jim\'enez-Vicente et al. 2012, 2014, Motta et al. 2012, Mosquera et al. 2013)
are large enough to mean that finite source sizes cannot be ignored.
Similarly,  quasar 
size estimates are also dependent on the fraction of mass in microlenses. 
Here, we carry out a joint analysis of both.
In Section 2 we describe the statistical analysis of the data based on microlensing simulations using magnification maps. In Section 3 we compare the results with previous studies and discuss the possible implications.

\section{Statistical analysis and results}

We use the microlensing magnification estimates for 27 quasar
image pairs in 19 lens systems from MED09. 
In order to have the largest possible sample but with a similar range of observed rest wavelengths, we include all objects 
from MED09 with magnifications
measured in the wavelength range between $\mathrm{Ly\alpha}$ (1216 \AA) and
Mg II (2798 \AA).With this choice, the average rest wavelength is
$\lambda = 1736\pm373$ \AA, while we
still keep 27 out of 29 image pairs from 19 out of 20 lensed quasars.
Only the system RXS J1131$-$1231 is excluded, as it was observed in [OIII], at a much larger wavelength of $\sim$5000 \AA. 
These microlensing magnification estimates are calculated after
subtracting the emission line flux ratios,  
which are little affected by microlensing (see e.g. Guerras et al. 2013),  
from the continuum flux ratios, and are therefore 
virtually free from extinction, substructure and macro
model effects (as these affect the line and continuum flux ratios equally).  
Our strategy is to compare
the observed microlensing magnification for a given image pair $\Delta m_i^{obs}$
with a statistical sample of simulated values for that measurement
as a function of the source size ($r_s$) and the fraction of
surface mass density in stars ($\alpha$). This will allow us to
calculate the likelihood of the
%observed microlensing magnifications given the parameters $P(\Delta m_i^{obs} | r_s,\alpha)$.
parameters $(r_s,\alpha)$ given the observations $L(r_s,\alpha|\Delta m_i^{obs})$.
The procedure is repeated for each of the 27 image pairs.
We calculate
magnification maps for each image using
a grid with
11 values for the fraction of the surface mass density in stars,  
$\alpha$, logarithmically distributed between 0.025 and 0.8 as 
$\alpha_j=0.025\times 2^{j/2}$ with $j=0,\cdots,10$. 
The 517 magnification maps were created
using the Inverse Polygon Mapping algorithm described by Mediavilla et al. 
(2006, 2011a). We used equal mass microlenses of $1 M_\sun$. All linear sizes
can be scaled for a different microlens mass as $\sqrt{M/M_\sun}$.
The maps have a size of 
$2000\times 2000$ pixels, with a pixel size of 0.5 light-days.
The maps therefore span 1000 light-days. The individual sizes of maps and pixels
in (more natural) units of Einstein radii for microlenses of $1M_\sun$ are given in Table \ref{tab1}.  On
average, the maps span
approximately 50 Einstein radii, with a pixel scale of roughly 0.025 Einstein radii.

\begin{deluxetable}{llrccc}
\tabletypesize{\footnotesize}
\tablewidth{0pt}
%\rotate
\tablecolumns{6}
\tablecaption{Microlensing data and some lens/map parameters.\label{tab1}}
\tablehead{
\colhead{Object} & \colhead{Pair} & \colhead{$\Delta m$} &
\colhead{$R_E/R_{eff}$} & \colhead{\shortstack{Map size \\ in $\eta_0$ }} & 
\colhead{\shortstack{Pixel size \\ in $\eta_0$ }}
}
\startdata 
HE0047$-$1756   & 	B$-$A   & -0.19	 & 1.63\tablenotemark{a} &  44.61 & 0.022 \\ 	
HE0435$-$1223   &	B$-$A   & -0.24	 & 1.60\tablenotemark{a} &  47.32 & 0.024 \\ 	
	        &	C$-$A   & -0.30	 & 			 &	  &	  \\	
	        &	D$-$A   &  0.09  &			 &	  &	  \\    
HE0512$-$3329   &	B$-$A   & -0.40  &			 &  79.08 & 0.039 \\
SDSS0806+2006   &	B$-$A   & -0.47	 & 3.30\tablenotemark{a} &  54.97 & 0.027 \\
SBS0909+532     &	B$-$A   & -0.60	 & 1.02\tablenotemark{b} &  77.88 & 0.039 \\ 
SDSS0924+0219   &	B$-$A   &  0.00	 & 2.93\tablenotemark{a} &  44.09 & 0.022 \\ 
FBQ0951+2635    &	B$-$A   & -0.69	 & 0.72\tablenotemark{b} &  35.61 & 0.018 \\ 
QSO0957+561     &	B$-$A   & -0.30	 & 1.29\tablenotemark{c} &  32.93 & 0.022 \\ 
SDSS1001+5027   &	B$-$A   &  0.23	 & 			 &  40.76 & 0.020 \\
SDSSJ1004+4112  & 	B$-$A   &  0.00	 & 			 &  59.11 & 0.030 \\
	        &	C$-$A   &  0.45  & 			 & 	  &	  \\
Q1017$-$20      &	B$-$A   & -0.26  & 1.46\tablenotemark{d} &  60.69 & 0.030 \\ 
HE1104$-$1805   &	B$-$A   &  0.60  & 2.19\tablenotemark{a} &  58.74 & 0.029 \\ 	
PG1115+080      &	A2$-$A1 & -0.65  & 2.48\tablenotemark{a} &  38.46 & 0.019 \\  	
SDSS1206+4332   & 	A$-$B	& -0.56  & 			 &  68.12 & 0.034 \\
SDSS1353+1138   &	A$-$B	&  0.00  & 			 &  38.03 & 0.019 \\
HE1413+117      &	B$-$A	&  0.00  & 			 &  61.16 & 0.034 \\
	        & 	C$-$A	& -0.25  & 			 &	  &	  \\  	
	        &	D$-$A	& -0.75  & 			 &	  &	  \\  	
BJ1422+231      &	A$-$B	&  0.16  & 2.29\tablenotemark{a} &  51.18 & 0.026 \\ 	
	        &	C$-$B	&  0.02  &			 &	  &	  \\	
	        &	D$-$B	& -0.08  & 			 &	  &	  \\  	
SBS1520+530     &	B$-$A	& -0.39  & 0.96\tablenotemark{b} &  60.15 & 0.030 \\
WFIJ2033$−$4723 &	B$-$C	& -0.50  & 1.56\tablenotemark{a} &  58.81 & 0.029 \\ 		
 	        &	A2$-$A1 &  0.00  &			 &	  &	  \\
\enddata
\tablecomments{ Microlensing Einstein radii $\eta_0$ used in columns 5 and 6 correspond to microlenses of $M=1M_\sun$.}
\tablenotetext{a}{From Oguri et al. (2014)}
\tablenotetext{b}{From Sluse et al. (2012)}
\tablenotetext{c}{From Fadely et al. (2010)}
\tablenotetext{d}{From Leh\'ar et al. (2000)}
\end{deluxetable}

The source size $r_s$ is taken into account by modelling
the source brightness profile as a Gaussian, $I(r)\propto \exp(-r^2/2r_s)$.
Mortonson et al. (1995) show that the specific shape of the radial profile 
is not important for microlensing studies because the results are essentially
controlled by the half-light radius rather than the detailed profile.
The Gaussian size $r_s$ is related 
to the half-light radius by $R_{1/2}=1.18 r_s$. 
To account for the source size, we convolve the 
magnification maps with Gaussians of 16 different sizes over a logarithmic
grid, $\ln(r_s/\mbox{lt-days})=0.3\times k$ with $k=0,\cdots,15$, which spans
$r_s\sim 1$ to $r_s\sim 90$ light-days. From the maps for a pair of images of a given
lensed quasar with fraction of stars $\alpha$ and convolved to size $r_s$,
we can calculate the likelihood of the parameters given the observed microlensing magnifications 
$\Delta m_i^{obs}$ as
\begin{equation}
L(r_s,\alpha|\Delta m_i^{obs})=P(\Delta m_i^{obs}|r_s,\alpha)\propto{\sum} e^{-\chi^2/2},
\label{eqlik}
\end{equation}
where
\begin{equation}
\chi^2=\frac{(\Delta m-\Delta m_i^{obs})^2}{\sigma^2}
\end{equation}
and $\sigma$ is a characteristic value for the error in the observed microlensing magnification (which we have set to 0.15 magnitudes). 
%$N$ is the number of trials with microlensing magnification $\Delta m$ for
%this set of parameters. 
For each image pair, the summation in Equation \ref{eqlik} is performed over $10^8$ trials 
by sampling the magnification map of each image at $10^4$ positions and summing over all possible combinations.
This procedure is repeated for the 176 possible values of the
$(r_s,\alpha)$ pairs, producing a 2D likelihood function 
for image pair $i$.
The process is repeated for each of the 27 pairs in our sample.
As we are using single epoch microlensing, the results for individual
pairs/objects 
have large uncertainties. 
\begin{figure}
\epsscale{0.85}
\plotone{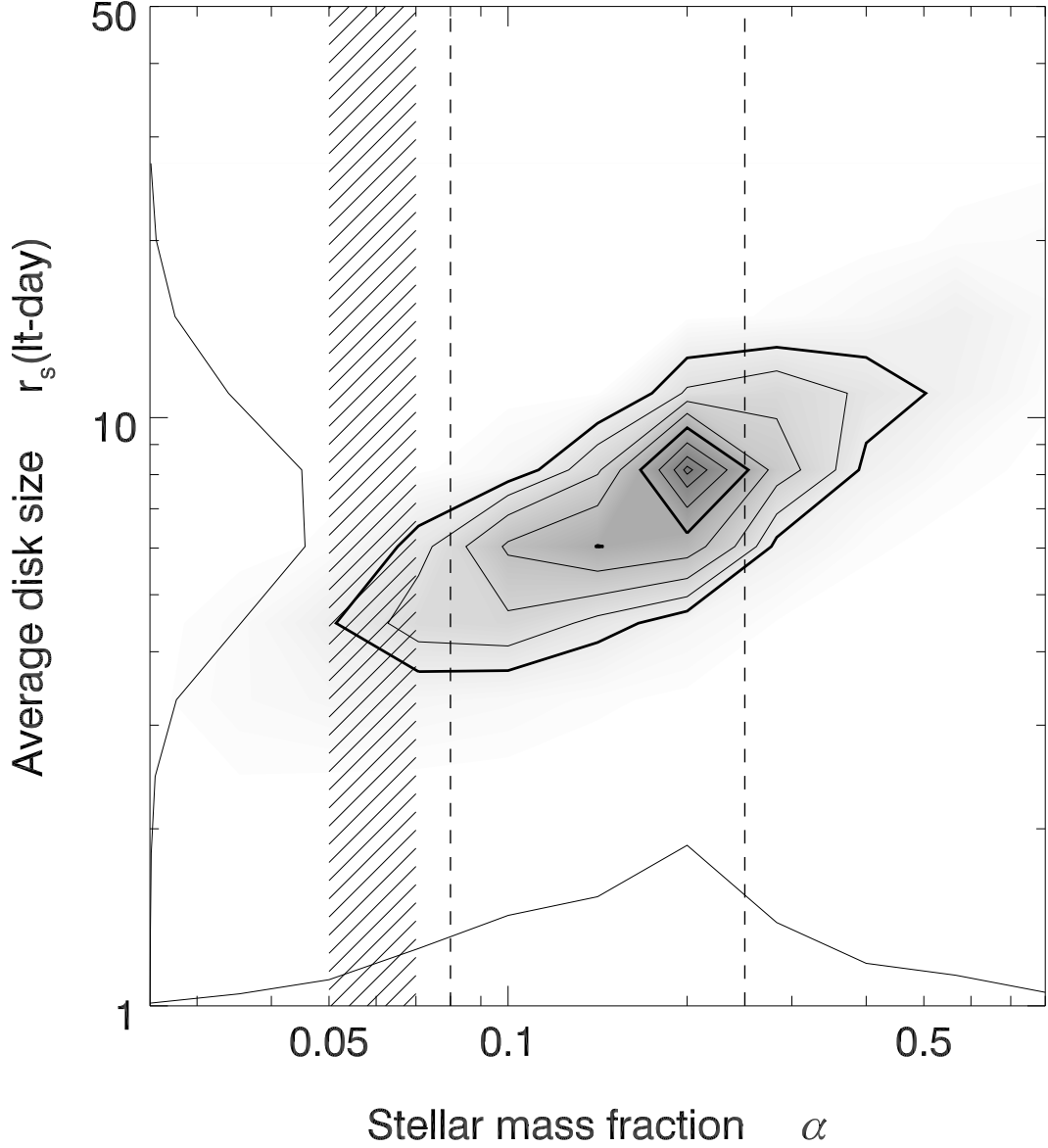}
\caption{Likelihood function for the fraction of mass $\alpha$ in the form of
stars or remnants, and the (Gaussian) size of the accretion disk $r_s$ at 1736 \AA (rest frame) for microlenses of mass $M=0.3M_\sun$. The contours are drawn at likelihood intervals of 0.25$\sigma$ for one parameter from the maximum. The contours at 1$\sigma$ and 2$\sigma$ are heavier.
The vertical striped band shows the previous estimates
from microlensing studies of lens samples by MED09 and Pooley
et al. (2012). The vertical dashed lines mark the region of estimates from microlensing studies of individual lenses (see text). 
The (marginalized) Bayesian posterior probability distributions using logarithmic priors for the
stellar mass fraction and disk size are shown along the corresponding axis. \label{fig1}}
\end{figure}

Since there is little signal in the individual pair likelihoods, we
combine the 27 likelihood distributions to produce a joint likelihood function
\begin{equation}
L(r_s,\alpha)\propto\prod_{i=1}^{27}L(r_s,\alpha|\Delta m_i^{obs}).
\end{equation}
The results of this procedure are shown in Figure \ref{fig1}.
The expected covariance between size and stellar mass fraction
found by MED09 can be clearly seen, but 
we find a well defined maximum in the likelihood distribution. 
The maximum likelihood estimate for the (average) mass fraction in stars is
$\alpha=0.2\pm0.1$ (at 68\% confidence level) and for the accretion disk size it is $r_s=8.1^{+4.1}_{-2.6}$ light days or, equivalently, $R_{1/2}=9.6^{+4.7}_{-3.2}$ light days
(for microlenses of 0.3$M_\sun$ and at a rest wavelength of roughly 1736 \AA). This value for the size of the
accretion disk is roughly 50\%-100\% larger than previously reported values 
but within the range of uncertainties (cf. Morgan et al. 2008, 2012, Blackburne et al. 2011, Mediavilla et al. 2011b, Mu\~noz et al. 2011, Jim\'enez-Vicente et al. 2012, 2014, Motta et al. 2012, Mosquera et al. 2013).

Figure \ref{fig1} also shows the posterior probabilities
for the two parameters in a Bayesian estimate with logarithmic priors on
the accretion disk size and the stellar mass fraction. In this case we have 
$P(r_s,\alpha)\propto P(r_s)P(\alpha)\prod_{i=1}^{27}L(r_s,\alpha|\Delta m_i^{obs})$ 
with $P(r_s)\propto 1/r_s$ and $P(\alpha)\propto 1/\alpha$. 
From these posterior probability distributions we
find Bayesian estimates of
$\alpha=0.21\pm 0.14$ at 68\% confidence, and  $r_s=6.7^{+3.3}_{-2.2}$ light days for microlenses of 0.3$M_\sun$, or $R_{1/2}=7.9^{+3.8}_{-2.6}$ light days, slightly smaller than the
maximum likelihood estimates.
If we increase the uncertainties to $\sigma=0.2$ mag, we find no significant changes in the estimates of the parameters. 
 We also recomputed the results sequentially dropping each lens and found that the results are not dominated by any single system.

\section{Discussion and conclusions}

\begin{figure}
\epsscale{0.85}
\plotone{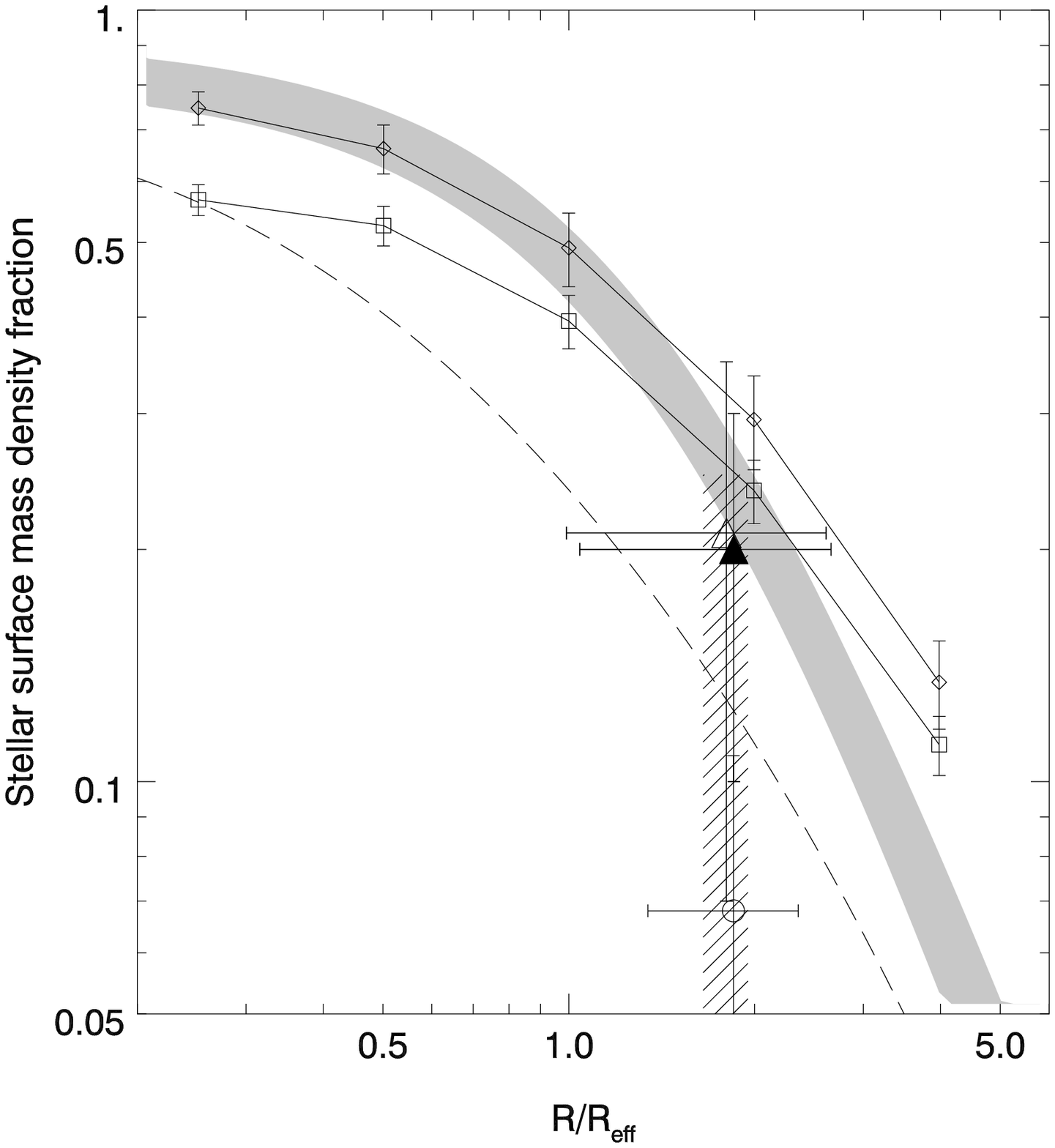}
\caption{Radial distribution of the stellar surface mass density fraction. 
The open (filled) triangle shows our maximum likelihood (Bayesian) estimate. The open circle is the estimate from Pooley et al. (2012). The vertical striped band shows the stellar mass fraction from MED09 for source sizes in the range between 0.3 (bottom) and 15.6 (top) light days.
The dashed line corresponds
to a simple model with a de Vaucouleurs stellar component and a total 
mass corresponding to a SIS with a flat rotation curve equal to the maximum rotational velocity of the stellar component. The grey band is the best fit profile for 
the sample of lenses analyzed by Oguri et al. (2014). The diamonds and squares correspond to a model using a Hernquist component for the stars, embedded in an NFW halo with (squares) and without (diamonds) adiabatic contraction of the dark matter, also from Oguri et al. (2014).\label{fig2}}
\end{figure}

With our joint analysis of stellar mass fraction and source size, we find a larger stellar
mass fraction than earlier statistical studies.
In Figure \ref{fig2} we compare our determination of the stellar
surface density fraction to a simple theoretical model and to the best fit to a sample of 
lens galaxies by Oguri et al. (2014).
The simple theoretical model is the early-type galaxy equivalent of a maximal disk model for spirals. We follow the rotation curve of a de Vaucouleurs component for the stars outwards
in radius until it reaches its maximum and then simply extend it as a flat rotation curve to become a singular isothermal sphere (SIS) at large radius (see details in Appendix \ref{ap1}). The ratio of the surface mass density of the de Vaucouleurs component to the total surface mass density is shown as a dashed curve in Figure \ref{fig2}.
We also show as a grey band 
the best fit for the stellar fraction in form of stars
determined by Oguri et al (2014) in a study of a large sample of lens galaxies
using strong lensing and photometry
as well as the best model using a
Hernquist component for the stars and an NFW halo for
the dark matter with and without adiabatic contraction also from Oguri et al. (2014).
We have used the average and dispersion estimates
for the Einstein and effective radii available for 13 of the objects in our sample from Oguri et al. (2014), Sluse et al. (2012), 
 Fadely et al. (2010) and Leh\'ar et al. (2000) (see Table \ref{tab1})
as an estimate of $R_E/R_{eff}$ in Figure \ref{fig2}. The average value and dispersion of the sample is $R_E/R_{eff}=1.8\pm0.8$.
This also averages over the different radii of the lensed images.
The agreement of our estimates with the expectations of the simple
theoretical model and with estimates from other studies (Oguri et al. 2014) 
is quite good. 
For comparison, the estimate of Pooley et al. (2012) (using the Einstein and effective 
radii estimates for 10 out of 14 of their objects from Schechter et al. 2014)
seems somewhat lower than expected at those radii. 
The range of stellar mass fractions from MED09 for source sizes in the range 0.3-15.6 light
days is also shown in Figure \ref{fig2}. In this case, the discrepancy between our estimate and
their reported value of $\alpha=0.05$ is completely due to the effect of the source size.
Although accretion disk sizes are known to be smaller in X-rays, recent estimates
are in the range 0.1-1 light-days, depending on the mass of the black hole
(cf. Mosquera et al. 2013), and 
these finite sizes will increase the stellar surface densitites implied by the X-ray data. 
Another possible origin for this discrepancy
is that Pooley et al. (2012) use the macro model as an unmicrolensed baseline for their 
analysis. It is well known that simple macro models are good at 
reproducing positions of images, but have difficulties reproducing the 
flux ratios of images due to a range of effects
beyond microlensing.
Recently, Schechter et al. (2014) found that 
the fundamental plane stellar mass densities have to be scaled up by a factor 1.52 in order 
to be compatible with microlensing in X-rays in a sample of lenses with a large 
overlap with that analyzed by Pooley et al. (2012). 
It is unclear how this need for more mass in 
stars at the position of the images found by Schechter
et al. (2014) can be reconciled with the apparently low estimate of mass in stars
at those radii by Pooley et al. (2012).
Our estimate of the stellar mass fraction agrees better 
with the results of 
microlensing studies of individual lenses (Keeton et al. 2006, Kochanek et al. 2006, 
Morgan et al. 2008, Chartas et al. 2009, Pooley et al. 2009, Dai et al. 2010b, Morgan et al. 2012) which reported values in the range 8-25\%, and with the estimates from strong lensing
studies (see for example Jiang \& Kochanek 2007, Gavazzi et al. 2007, Treu 2010, Auger et al. 2010, Treu et al. 2010, Leier et al. 2011, Oguri et al. 2014) which produced stellar mass
fractions in the range 30-70\% integrated inside the Einstein radius of the lenses.  

The estimated size of the accretion disk in this work is 
slightly larger than the results found by other authors but still compatible with them
(cf. Morgan et al. 2008, Blackburne et al. 2011, Mediavilla et al. 2011b, Mu\~noz et al. 2011, Jim\'enez-Vicente et al. 2012, 2014, Motta et al. 2012, Mosquera et al. 2013). Those studies find values roughly in the range of 4-5 light-days. 
Thus, our present estimate for the size of the accretion disk, 
mantains the discrepancy with the 
simple thin disk model (Shakura \& Sunyaev, 1973) 
that predicts accretion disks of sizes roughly 2-3 times smaller.  
Spectroscopy (preferably
at several epochs) for a larger sample of lens systems would 
allow us to expand the sample
and to extend its conclusions. 
A larger sample could also be divided 
into statistically significant suitable subsamples, 
to examine the dependence of the stellar mass fraction on radius, 
lens mass or redshift, 
questions which are difficult to probe by other means.

\acknowledgements

 The authors would like to thank M. Oguri for kindly providing the differential version of their results for comparison with the present work shown in Figure 2. 
We are also grateful to the anonymous referee for useful suggestions that improved the presentation of this work.
This research was supported by the Spanish Mi\-nis\-te\-rio de Educaci\'{o}n y Ciencia with the grants AYA2011-24728, AYA2010-21741-C03-01 and AYA2010-21741-C03-02. JJV is also supported by the Junta de Andaluc\'{\i}a
  through the FQM-108 project. JAM is also supported by the Generalitat Valenciana with the grant PROMETEOII/2014/060.
CSK is supported by NSF grant AST-1009756.

\appendix
\section{A simple theoretical model for the local stellar mass fraction in lens galaxies}
\label{ap1}

We describe here the simple theoretical model shown as a dashed curve in Figure \ref{fig2}. In this model, the stars are distributed according to a de Vaucouleurs model. We will use units of the effective radius
of this de Vaucouleurs system so that $x=r/R_{eff}$. 
The surface density of stars is given by the de Vaucouleurs law:
\begin{equation}
\Sigma_s(x)=\Sigma_{0} e^{-k\ x^{1/4}},
\end{equation}
with $k=7.66925001$. $\Sigma_{0}$ is the stellar surface mass density at the galaxy center. The mass enclosed within radius $x$ in this system is given by:
\begin{equation}
M_s(x) = \frac{\Sigma_{0}40320\pi}{k^8}\left[1.0-e^{-q}\left(1+q+\frac{q^2}{2!}+\frac{q^3}{3!}+\frac{q^4}{4!}+\frac{q^5}{5!}+\frac{q^6}{6!}+\frac{q^7}{7!}\right)\right],
\end{equation}
where $q=kx^{1/4}$ (cf. Maoz \& Rix, 1993). The rotation curve of this stellar system (in units in which $G=1$) is given by $v_s(x)=\sqrt{M_s(x)/x}$.
Let $v_{\rm max}$ be the maximum of the rotation curve of this stellar system. We model the total mass of the lens galaxy as an SIS with a (flat) rotation curve equal to that $v_{\rm max}$. The total mass of the system has therefore a surface mass density given by:
\begin{equation}
\Sigma_t(x)=\frac{v_{\rm max}^2}{4 x},
\end{equation}
and the (local) fraction of mass in form of stars, which is represented as a dashed line in Figure \ref{fig2} is the ratio of the stellar to total mass surface densities, $\alpha(x)=\Sigma_s(x)/\Sigma_t(x)$. Note that $\Sigma_t \propto v_{\rm max}^2 \propto \Sigma_0$ so that $\alpha(x)$ is independent of the value of $\Sigma_0$.

\end{document}